\documentclass[article,10pt]{article}
\usepackage[utf8]{inputenc}
\usepackage{graphicx}
\usepackage{color}
\usepackage{amsmath}
\usepackage{authblk}
\title{Real Space Migdal-Kadanoff Renormalisation of Glassy Systems: Recent Results and a Critical Assessment}
\author[1]{Maria Chiara Angelini}
\author[2,3]{Giulio Biroli}
\affil[1]{Dipartimento di Fisica, Ed. Marconi, "Sapienza" Universit\`a di Roma, P.le A. Moro 2, 00185 Roma Italy}
\affil[2]{Institut Physique Th\'eorique (IPhT) CEA Saclay, and CNRS URA 2306, 91191 Gif Sur Yvette, }
\affil[3]{Laboratoire de Physique Statistique, Ecole Normale Sup\'erieure, 24 rue Lhomond, 75005 Paris, France.}
\date{}

\def\<{\langle}
\def\>{\rangle}
\def\(({\left(}
\def\)){\right)}
\def\[[{\left[}
\def\]]{\right]}

\begin{document}

\maketitle

\begin{abstract}
In this manuscript, in honour of L. Kadanoff, we present recent progress obtained in the description of finite dimensional glassy systems thanks to the Migdal-Kadanoff renormalisation group (MK-RG). We provide a critical assessment of the method, in particular discuss its limitation in describing situations in which an infinite number of pure states might be present, and analyse the MK-RG flow in the limit of infinite dimensions. 
MK-RG predicts that the spin-glass transition in a field and the glass transition are governed by zero-temperature fixed points of the renormalization group flow. This implies a typical energy scale that grows, approaching the transition, as a power of the correlation length, thus leading to enormously large time-scales as expected from experiments and simulations. These fixed points exist only in dimensions larger than $d_L>3$ but they nevertheless influence the RG flow below it, in particular in three dimensions. MK-RG thus predicts a similar behavior for spin-glasses in a field and models of glasses  and relates it to the presence of avoided critical points.     
\end{abstract}

\section{Introduction}\label{Sec:intro}
In the last thirty years the field of disordered systems was a remarkable fertile ground. 
In the struggle to understand the physics of disordered systems physicists developed 
new ideas and new tools, whose relevance actually goes beyond physics 
itself.  Yet, despite a lot of progress, a finite dimensional theory of archetypical disordered systems like spin-glasses and glasses is still lacking. It is not for lack of imagination, indeed
several theories have been proposed, among which there are very solid and deep ones \cite{spin-glass beyond,droplet,dropletBM,RFOT,Wbook,GarrahanChandlerreview}. \\
One crucial missing piece is clearly a renormalization group approach able to cope with the complexity 
of these problems. Mean-field theory \cite{FRSB,RFOT} unveiled that 
the order parameter for these transitions is a complicated abstract object, the so-called 
Parisi matrix that was first introduced in studies of spin-glasses \cite{FRSB}. Developing 
a field theoretical renormalization group (RG) procedure able to include and asses the role of fluctuations on top of mean-field theory has been proved to be a formidable challenge yet to be solved. 
The two main difficulties besides the intricate nature of the order parameter is that the RG has to be functional and non-perturbative to be able to capture the  complexity of the 
problem---certainly not an easy task!\\
In this context the real space RG methods pioneered in the mid 70s by L. Kadanoff \cite{Kadanoff} provide a very useful way out from this theoretical impasse. 
They allow to integrate out short-scale degrees of freedom and replace 
the original disordered system with a new one characterized by renormalized 
couplings. These methods are approximate, higher order couplings created 
by the RG procedure are neglected without any real justification, but 
they are able to provide remarkable predictions, in particular they can address problems 
where the RG has to be functional and non-perturbative as it is the case for glassy systems. \\
For instance in the context of the Random Field Ising Model they correctly capture the nature of the critical point: 
a zero temperature fixed point for which a field theoretical RG treatment was established only very recently \cite{tarjustissier}. 
In the context of spin-glasses they are at the basis of the so-called droplet theory which is one of the two competing scenario for describing the physics 
of finite dimensional systems \cite{droplet,dropletBM}
(the other being Parisi's mean-field theory). \\
The real space RG, in particular the version introduced by Migdal and Kadanoff (MK) \cite{Migdal} has other advantages, 
in particular it is exact in one dimension and often provides a good quantitative approximation for the values of the critical exponents in not too high dimensions.  
All in all, despite their approximate and uncontrolled nature, real space RG methods have proven 
 to provide valuable guidelines for the behaviour of finite dimensional systems. 
 They have a predictive power for RG flows similar to the one of mean-field theory for phase diagrams: 
 they provide a qualitative correct description of RG flows, very often describing correctly the nature of the fixed points but being unable to provide accurate results except in low dimensions. \\
In this work, in honour of L. Kadanoff, we present the recent progress obtained in the description of finite dimensional glassy systems thanks to the MK-RG method and provide 
a critical assessment. 

\section{The Migdal-Kadanoff Renormalisation Group Method for disordered systems}
A reliable RG method for disordered systems should be able to follow the flow of the whole distribution of couplings and fields. This is why it has to be functional, which is indeed 
the case for the MK-RG procedure.

\begin{figure}[t]\label{Fig:MK}
 \includegraphics[width=\textwidth]{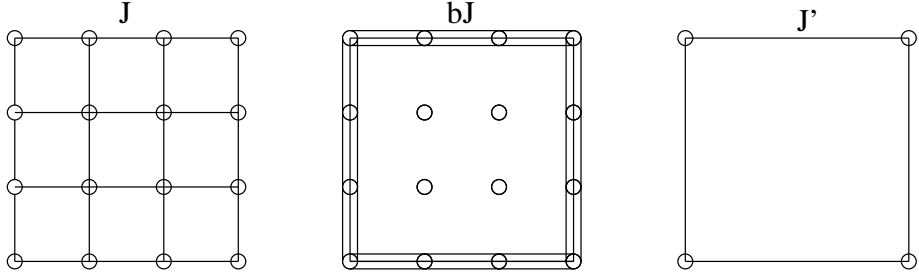}
\caption{Example of MK RG on a two-dimensional square lattice. The block-size is $b=3$ in this case.}
 \end{figure}

\begin{figure}[t]\label{Fig:HL}
 \includegraphics[width=\textwidth]{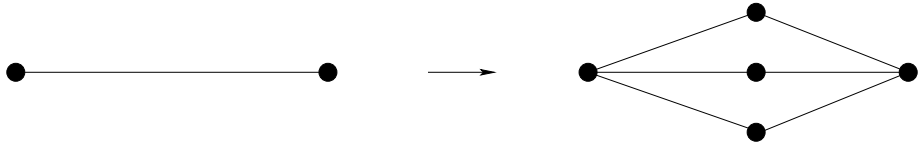}
\caption{Basic step for the construction of a hierarchical lattice with parameters $p=3$ and $s=2$. }
 \end{figure}

The MK method, which is a type of approximated Real Space (RS) transformation, has 
several advantages in the context of disordered systems. It is physically very transparent, unlike 
field theoretical analysis, or $\epsilon$-expansion, of replica field theory. 
Moreover it is non-perturbative: it has the potentiality of capturing non-perturbative effects that are known to be important for
disordered systems in finite dimensions.
The main drawback is that RSRG can not be carried out exactly in more than one dimension because higher order couplings between any subset of spins are created. For this reason many approximations have been developed. These approximations do not leave the partition function invariant, as required by exact RG, however they can provide a good approximation of the RG flow.
Among these approximations, there are the lower bound transformations, introduced
by Kadanoff in Ref. \cite{Kadanoff}, which consists in replacing the Hamiltonian of the system,$H(\sigma)$, by $H(\sigma) + V (\sigma)$ where $V(\sigma)$ is chosen so that the sum over
the spins configurations can be evaluated explicitly.
If the chosen potential has the special property $< V >_H = 0$, the free energy of the renormalised
system is a lower bound to the free energy of the original system. For the nearest
neighbor Ising model, one good choice for $V(\sigma)$ is the bond moving potential. 
The property $< V >_H = 0$ is satisfied thanks
to the translational invariance. The effect of this kind of potential is to move some
couplings from two spins to other ones, in such a way that some spins can be
decoupled. This is the basis of the Migdal-Kadanoff renormalization \cite{Migdal}.\\
Let us see how it practically works on a nearest
neighbour Ising model (a 2-dimensional example is shown in Fig. \ref{Fig:MK}). The
spins are divided in blocks of size $b$. All the internal couplings are moved to the spins
at the edges of the blocks, their sum being $b^{d-1}\cdot J$. At this point a decimation (a partial sum) of the
spins at the edges except that on the corners is performed obtaining $J' = \mathcal{R}_b(b^{d-1}J)$,
where the function $\mathcal{R}_b(J)$ is the one that enters in the exact renormalisation of a one-dimensional system. 
The final result will depend of our
choice of b, because we are not making an exact RG procedure. The best choice
is to focus on infinitesimal transformations. 
 In two dimensions the fixed point of the infinitesimal transformation 
for the ferromagnetic Ising model, that gives the critical temperature, is the correct one
thanks to the fact that the 2D square lattice is self dual.\\
Berker and Ostlund realized that for some particular lattices, the hierarchical diamond lattices, the
MK renormalization procedure is exact \cite{Berker}. These lattices can be generated iteratively as in Fig. \ref{Fig:HL}. 
The procedure starts at the step $G = 0$ with two spins connected by a single link. At each step $G$, the
construction is applied to each link of step $G-1$. For each link, $p$ parallel
branches, made of series of $s$ bonds, are added, with $p \cdot (s - 1)$ new spins. The
effective dimension of this model is $d = 1 + ln(p)/ ln(s)$. In fact in a standard $d$
dimensional lattice, if the length grows of a factor $L$, the number of links grows
with a factor $L\cdot d$ . If in the hierarchical lattice the length grows of a factor $s$, the
number of links grows of a factor $p \cdot s$. However the parameter $d$
is not sufficient to identify the lattice, because the same $d$ can be obtained with different $s$ and $p$,
and the renormalisation will produce different results (as in the MK RG on a hyper-cubic lattice for different $b$). 
The renormalisation of these lattices can be performed exactly, first tracing (summing) on the $s-1$ internal spins on each branch
and then from a lattice at the G generation, and then summing on the $p$ branches at each step.

This property continues to hold in presence of quenched disorder: the MK renormalisation remains exact on hierarchical lattices.
In this case, however, for each disorder sample in the original lattice, 
one ends up with different renormalised couplings after one RG step: instead of following the renormalization of a few couplings and fields, one has to focus on 
the flow of their disorder probability distribution. 
%Usually, it is sufficient to look at simple parameters as mean and variance
%of this distribution to identify the transitions.

Among the successes of the MK-RG for disordered systems stands the Random Field Ising Model, for which the MK renormalization correctly captures the nature of the transition and of the corresponding FP of the RG flow (a zero temperature FP)
\cite{MKrfim}. Another one is the correct identification of the nature of the FP and the existence of a phase transition for spin glasses
(in zero magnetic field), as well as the value of the lower critical dimension, $d_L=2.5$, the same one obtained in 
numerical analysis \cite{dLboettcher}.
All these are remarkable features but there are downsides too. In particular, hierarchical lattices are not able to recover the infinite dimensional solution corresponding to mean-field theory.
This drawback, already known for non-disordered systems (the existence of an upper
critical dimension is missed), is even more serious in the case of disordered systems. 
As shown in Ref. \cite{Gardner}, spin glasses
on hierarchical lattices are "replica symmetric" even in the infinite dimensional limit. 
This means that the MK approximation for hyper-cubic lattices is 
not able to capture the complex mean-field theory developed by Parisi and rigorously known 
to be exact for infinite dimensional lattices. When interpreting MK results for glasses and spin-glasses is important to keep this point in mind and ask whether the results one finds 
are in contrast with mean-field theory predictions or, instead, MK is reproducing in a very crude
way a more complex scenario related to replica symmetry breaking. We will come back to this point later on. Other discussions on the limitations of MK RG in the case of disordered systems can be found in \cite{Antenucci}.\\

\section{Models and methods}\label{Sec:MKRG}
We now introduce the models of glasses and spin-glasses that have been analyzed using MK-RG.  The MK RG provides new insights and predictions for both of them. 

There are two classes of models:
\begin{itemize}
 \item \textbf{The Edwards-Anderson model of spin glasses}:
 the variables are Ising spins taking the values $\sigma=\pm 1$ and the Hamiltonian is:
 \begin{equation}
 H(\{\sigma\})=-\frac{1}{p}\((\sum_{\langle i,j \rangle}J_{ij}\sigma_i\sigma_j+\sum_i\sigma_i h_i\)),
 \label{eq:H_SGH}
\end{equation}
where $J_{ij}$ and $h_i$ are independent random variables extracted from a Gaussian distribution (other distributions can also be considered). We denote  the variance of the couplings and the fields $V_J^2$ and $V_H^2$ respectively ($V_J=1$ in the following).
The factor $\frac{1}{p}$, which is related to the space dimension $d$ as $d = 1 + ln(p)/ ln(2)$, allows to have a good $d\rightarrow\infty$ limit for the MK-RG method. 
 
 \item \textbf{Disordered models of glasses: the $M$-value models.} 
Several disordered models of glasses have been introduced in the literature: 
Random Energy Model (REM) \cite{derrida,grossmezard}, the p-spin models, etc. 
They are all at the basis of the Random First Order Transition theory \cite{RFOT}. In particular,
their mean-field solution displays a glassy 1step Replica Symmetry Breaking (1RSB) and 
an associated entropy crisis transition \`a la Kauzmann. The finite dimensional versions
of all these systems can be described within the same class of models, that we called M-value models and are defined as follows. \\
Consider variables ("spins") that can take $q=2^M$ values and a Hamiltonian of the form:
\begin{equation}
H(\{\sigma\})=\sum_{\langle i,j \rangle}E_{i,j}(\sigma_i, \sigma_j).
\label{eq:H_Mvalue}
\end{equation}
where $E_{i,j}(\sigma_i, \sigma_j)$ are a set of Gaussian random link-energies. The different models studies in the literature can be realized by choosing appropriately the covariance of the $E_{i,j}(\sigma_i, \sigma_j)$s. 
The simplest case is the finite-dimensional version of the Random Energy Model (REM) \cite{derrida,grossmezard,REM1D}, in which $E_{i,j}(\sigma_i, \sigma_j)$ are just
independent random variables extracted from a Gaussian distribution. 
The $M-p$ spins disordered models \cite{Mp}, finite-dimensional versions of the p-spin disordered models, as well as Potts \cite{sompo} and Super-Potts (SP) glasses \cite{SP} can also be reproduced. \\
%These are finite-dimensional versions of the p-spin disordered models, extensively studied as the archetype for the 1RSB transition and for mean-field 
%models of glasses. The $M-p$ model has an RFOT transition in its mean-field version for large enough values of $M>M_c(p)$ \cite{Mp}. 
%Also Potts-models and Potts-glass models are easily shown to belong to the $M$-value class, as well as the Super-Potts (SP) Model.
Let's consider for instance Super-Potts models: on each site there is a Potts variable characterised by $q=2^M$ "colors" and 
 $ E_{i,j}(\sigma_i, \sigma_j)=E_0$ for $(\sigma_i, \sigma_j)=(\sigma_i^*, \sigma_j^*)$ and 
$ E_{i,j}(\sigma_i, \sigma_j)=E_1$ otherwise; $(\sigma_i^*, \sigma_j^*)$ is randomly drawn among the 
$2^M\times 2^M$ possible couples $(\sigma_i,\sigma_j)$ independently for any couple of neighbors $(i,j)$ \cite{SP}. These systems have been shown analytically 
to display an RFOT transition within mean-field theory for $2^M$ larger than $20$ and numerically to display evidences of a glass transition in three dimensions for $2^M$ larger than $30$ \cite{SP}.
Super-Potts models are generalisations of the disordered Potts glasses \cite{sompo} 
that originally inspired RFOT theory \cite{RFOT} \footnote{Unfortunately disordered nearest neighbour Potts glasses do not display glassy 
phenomenology in three dimension. For this reason we introduced SP model in order to bypass the problems (weak frustration and lack of glassy behaviour) 
found for disordered Potts model.}.
REM, $M-p$ and SP model were explicitly analyzed through MK RG in ref. \cite{MKREM}.
Also the random permutation Potts glasses \cite{PermutationMarinari1,PermutationMarinari2} and the third nearest neighbours 
disordered Potts model recently introduced \cite{giap} belong to the $M$-value class and could be studied as well. 
Note that the MK RG that we use requires the interactions between the $\sigma_i$ of the closest blocks to be pairwise, as they are in the $M$-value models.
However one can also consider models with multi-$\sigma_i$ finite range interactions.  
Since the interaction range is finite, i.e. degrees of freedom distant more than $\ell_{I}$ do not interact, one can divide the lattice in a sub-lattice of blocks of size $\ell_{I}^d$. The degrees of freedom inside one block can be "packed" into one single degree of freedom $\sigma_i$ that takes $q=2^{M'}$ values, where e.g. ${M'}={\ell_{I}^d}$  for spin  variables. In terms of these new variables $\sigma_i$ one obtains again a model with interaction between closest neighbours only \footnote{Residual multi-$\sigma_i$ interactions between closest blocks, if present, can be safely neglected remain since they become negligible with respect to the pairwise interaction between nearest neighbour blocks 
if the size of the blocks is taken much larger than $\ell_{I}$.} . 
In this way even multi-$\sigma_i$ interactions models can be placed in the class of $M$-value models.
\end{itemize}

The Edwards-Anderson model is a particular case of $M$-value models for which $M=1$ and the energies are chosen following eq. (\ref{eq:H_SGH}).
It has to be considered separately for two main reasons: first, is a model of spin-glasses and not glasses, i.e. its phase diagram (as well as its mean-field solution) is different; second, 
in absence of the field it displays the up-down spin symmetry which leads to important
consequence for the RG flow, as we shall discuss. 

Let us now present the MK-RG procedure in detail. For both models in eqs. (\ref{eq:H_SGH}) and (\ref{eq:H_Mvalue}), the renormalisation steps can be described in the same way. The main observation in order to do so is that we can always decompose the energy $E_{i,j}$ in coupling-like contributions, fields-like contributions plus a constant:
\begin{equation}
E_{i,j}(\sigma_i,\sigma_j)=-J(\sigma_i,\sigma_j)-H_L(\sigma_i)-H_R(\sigma_j)+C
\label{eq:decompositionE}
\end{equation}
where $J$, $H$ and $C$ are identified by the relations:
$$C=\frac{1}{2^{2M}}\sum_{\sigma_i,\sigma_j}E(\sigma_i,\sigma_j);$$

$$
 H_R(\sigma_j)=-\frac{1}{2^M}\sum_{\sigma_i} E(\sigma_i,\sigma_j)+C \,;\, H_L(\sigma_i)=-\frac{1}{2^M}\sum_{\sigma_j} E(\sigma_i,\sigma_j)+C
 $$
  $$
 J(\sigma_i,\sigma_j)=-E(\sigma_i,\sigma_j)+\frac{1}{2^M}\sum_{\sigma_i} E(\sigma_i,\sigma_j)+\frac{1}{2^M}\sum_{\sigma_j} E(\sigma_i,\sigma_j)-C
 $$
 and have the following properties:
\begin{equation}
\sum_{\sigma_i}J(\sigma_i,\sigma_j)=\sum_{\sigma_j}J(\sigma_i,\sigma_j)=\sum_{\sigma_i}H_L(\sigma_i)=\sum_{\sigma_j}H_R(\sigma_j)=0
\label{eq:propJH}
\end{equation}
 
 It is easy to verify that for the SG in field these equations for a single link lead exactly to $h_i$, $h_j$ and $J_{ij}$ ($C=0$). 
 Note that even if some of those terms are not present in the original model, they can be generated by the 
 RG flow \footnote{Let's just point out a subtlety: the field presents in eq. (\ref{eq:H_SGH}) are site-fields while the ones present in eq. (\ref{eq:H_Mvalue}) are link-fields, i.e. associated to a link. The renormalisation of eq. (\ref{eq:H_SGH}) can produce additional link-fields, in addition to the site ones, exactly as happens for the M-value models.}. \\
 Let's now consider the spin-glass and glass models defined previously on a hierarchical lattice with parameter $s=2$, whereas the parameter $p$ can be varied to tune the effective dimension.
 The renormalisation on the hierarchical lattice is the inverse operation with respect to its creation, described in fig. \ref{Fig:HL}, 
 and can be decomposed in two main steps: firstly we perform the sum on the internal spins, that we will call $\sigma_3^i$, on each of the $p$ branches. 
 This step generates $p$ interactions between the
 external spins, namely $\sigma_1$ and $\sigma_2$. In the second step we sum the contribution of these $p$ different interactions.
 The new renormalised energy $E^R_{1,2}$ between the two external spins reads:
\begin{equation}
e^{-\beta E^R_{1,2}(\sigma_1,\sigma_2)}\equiv \prod_{i=1}^p\left(\sum_{\sigma_3^i=1}^{2^M}e^{-\beta \left((E_{1,3}^i(\sigma_1,\sigma_3^i)+E_{3,2}^i(\sigma_3^i,\sigma_2)+h_3^i\sigma_3^i\right))}\right)
\label{eq:RG1}
\end{equation}
where we have put explicitly the site-fields $h_3^i$ that can be present in the SG in field in addition to the link-fields already included in $E_{1,3}$ and $E_{3,2}$.
The renormalised energy $E^R_{1,2}$  is again of the type in eq. (\ref{eq:decompositionE}).
In this way after one RG-step we get a renormalised model whose unit of length is doubled.    
Eq. (\ref{eq:RG1}) gives the relation between the new renormalised couplings and fields as a function of the old ones. Since it is an identity between random variables, it provides the flow equation for their distribution, i.e. the distribution of the quenched disorder.

Finally, let us stress that in the presence of external fields there is a difference between HL and
bond-moving MK on an hyper-cubic lattice. We follow ref. \cite{MKSGHMoore} and move
the fields coherently with the bonds on the spins placed on
the edges of the blocks that are traced out in the RG
step. In this way, the RG iteration is almost exactly the same
one of a HL. At each RG-step a new coupling and two new sites fields 
are generated. The difference with usual RG on HL is that these sites fields   
are moved from the external spins to the internal
ones for all $p$ branches but one. The unmoved fields represent the ones on the original link. None of the original
site-fields is moved. This change in the renormalisation
procedure is important to have a correct interpretation
in terms of bond moving and to avoid pathological behaviours. 
If the fields are not moved from the external spins to the internal ones, even in the ferromagnetic model with external field 
one would get a renormalised field that grows under renormalisation even when the coupling goes to zero at the paramagnetic fixed point. The field moving takes care of this problem
ensuring that the field remains constant once the couplings have gone to zero.
 
\section{Spin Glass models} 

In this section we will first review the state of the art, present the results obtained for the spin-glass in a field by MK renormalisation and discuss the new perspectives they offer. 
\subsection{State of the art}
In order to explain the physics of SGs, two main theories were developed: the Full Replica Symmetry Breaking (FRSB) and the Droplet Theory (DT). The former is based on the exact solution of mean-field models, whereas the latter is based on a low temperature scaling theory supported by MK-RG.  
Concerning finite dimensional SGs in zero magnetic field, the FRSB theory claims that the scenario remains 
to large extent the same as the MF one, in particular  there are many competing pure states that dominate the free energy in the low temperature phase \cite{FRSB}.
Contrary, DT assumes that the pure states are just two, related by the inversion symmetry \cite{droplet,dropletBM}.
The prediction of the two theories are very different, nonetheless numerical simulations and experiments are not able to clearly affirm the exactness of one or the other,
the large finite size effects and equilibration times being the main obstacles \cite{review}.
The differences in the predictions of FRSB against DT become even more drastic when an external field (random or constant) is added:
in the FRSB scenario if the field is sufficiently small, the number of the pure states will remain large: a small field is not able to destroy the SG transition that persists below 
a certain line, the so called De Almeida-Thouless (AT) line, in the field-temperature plane \cite{AT}.
Contrary in the DT scenario the external field will select just one of the two pure states: the transition disappears adding an infinitesimal field. Given the very different scenari predicted in this case, much of the research has recently focused on SGs in a field with the hope of being able 
to identify the correct one. Unfortunately, numerical results are still not clear cut enough to identify the correct theory even in this case \cite{SGHnum1,SGHnum2,SGHnum3}.

Given the difficulties encountered in numerical simulations, as usual in the study of phase transitions, researchers tried from the very beginning to tackle the problem using RG. Also this, however, turned out to be quite intricate.  
The field theoretical analysis showed that the Gaussian fixed point (FP) that controls the critical behaviour of the AT line for MF model becomes unstable for $d < 6$ \cite{BrayRoberts}. No perturbative fixed point exists below six dimensions, in consequence the usual strategy of performing an $\epsilon$-expansion around the MF critical fixed-point is useless. 
For $d>6$ , the basin of attraction of the Gaussian FP is finite, and shrinks to zero at $d=6$ \cite{FiniteBasin}.
These results do not necessarily means that there is no transition below six dimension, but 
if there is one then it has to be associated to non-perturbative fixed point (NP-FP). 
Note that this putative NP-FP, since it lies outside the basin of attraction of the Gaussian one, 
can govern the physics of the system even above 6 dimensions, it depends in which basin
of attraction the initial condition of the RG flow, corresponding to finite dimensional SGs, lies. 
This means that the upper critical dimension above which MF behavior is recovered is expected to be larger than six, possibly much higher ($d=\infty$ is also a possibility).  
Very recent results obtained by two loop expansions and NP-RG \`a la Wetterich suggest 
the existence a stable non-perturbative FP but only below $d=5.4$ thus providing a new interesting facet to this intricate situation \cite{Yaida}. \\
Given this state of  the art, and the necessity of looking for NP-FP, a thorough analysis by 
MK-RG is very valuable. As we show below, it provides an interesting new perspective on the problem. Part of the results we present in the following have been published in \cite{MKSGH}. 

\subsection{The MK RG for the Spin Glass in a field}
We now present the application of MK-RG to the problem of SGs in a field. The main result
we shall find is that a phase transition, related to a NP-FP, is found for $d\ge d_L=8.066$ \cite{MKSGH} (no transition 
was found for $d=2,3,4$ with the same method in ref. \cite{MKSGHMoore}).

\begin{figure}[t]
 \includegraphics[width=\textwidth]{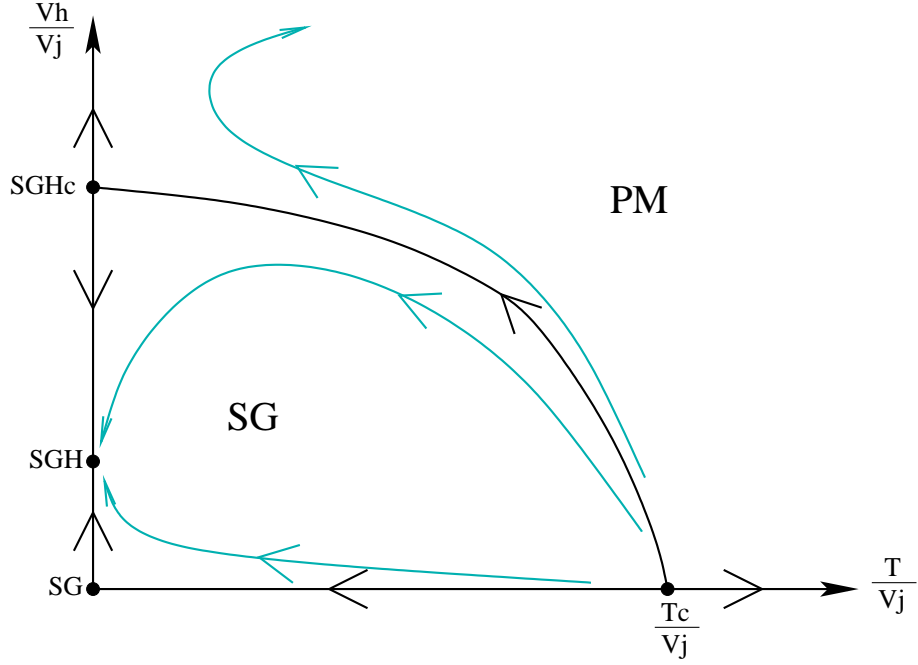}
\caption{Qualitative RG diagram for the SG in field at $d\ge 8.066$. Three zero temperature FP are present, the critical one, the zero-field one and the SG in field one (from ref. \cite{MKSGH}).}
\label{Fig:diagramSGH}
 \end{figure}

The renormalisation eq. (\ref{eq:RG1}), with the movement of the fields for the model with Hamiltonian (\ref{eq:H_SGH}) explicitly reads
\begin{align}
\nonumber
&-\frac{\beta}{p} E^R_{1,2}(\sigma_1,\sigma_2)=\\
=&\ln\[[\sum_{\sigma_3^p=\pm1}
\nonumber
e^{-\frac{\beta}{p} \((J_{1,3}^p(\sigma_1,\sigma_3^p)+\overrightarrow{h^p}_1\sigma_1+\overleftarrow{h^p}_3\sigma_3^p+
J_{3,2}^p(\sigma_3^p,\sigma_2)+\overrightarrow{h^p}_3\sigma_3^p+\overleftarrow{h^p}_2\sigma_2+h_3^p\sigma_3^p\))}\]]+\\
&+\sum_{i=1}^{p-1}\ln\[[\sum_{\sigma_3^i=\pm1}
e^{-\frac{\beta}{p} \((J_{1,3}^i(\sigma_1,\sigma_3^i)+(\overrightarrow{h^i}_1+\overleftarrow{h^i}_3)\sigma_3^i+
J_{3,2}^p(\sigma_3^i,\sigma_2)+(\overrightarrow{h^i}_3+\overleftarrow{h^i}_2)\sigma_3^i+h_3^i\sigma_3^i\))}\]]
\label{Eq:RGiterSGH}
\end{align}
where we have called the generated link fields $\overrightarrow{h}$ and $\overleftarrow{h}$, while $h$s are the original site-fields.
We chose the $p$-th branch as the original one, for which link-fields are not moved. Site fields are never moved.
Eq. (\ref{Eq:RGiterSGH}) leads to different equations for the renormalised couplings and fields:
\begin{align}
J^R_{12}=&\frac{p}{4\beta}\[[R(1,1)+R(-1,-1)-R(1,-1)-R(-1,1)\]]
\label{eq:Jn_SGH}
\end{align}

\begin{align}
\overrightarrow{h}^R_1=&\overrightarrow{h}_1^p+\frac{p}{4\beta}\[[R(1,1)+R(1,-1)-R(-1,1)-R(-1,-1)\]]
\label{eq:Hn_SGH}
\end{align}
with
\begin{align}
R(\sigma_1,\sigma_2)=\sum_{i=1}^{p}\log\((\sum_{\sigma_3^i=\pm1}e^{-\frac{\beta}{i} \((J_{1,3}^i(\sigma_1,\sigma_3^i)+H^i\sigma_3^i+
J_{3,2}^i(\sigma_3^i,\sigma_2)\))}\))
\end{align}
where we defined $H^p=\overleftarrow{h^p}_3
+\overrightarrow{h^p}_3+h_3^p$ and $H^i=\overrightarrow{h^i}_1+\overleftarrow{h^i}_3
+\overrightarrow{h^i}_3+\overleftarrow{h^i}_2+h_3^i$ for $i=(1,...,p-1)$.
These equations define the RG-flow of the probability distributions of the disorder. They are solved by the population dynamics method \cite{popdyn}. \\
The results are summarised in fig. \ref{Fig:diagramSGH} by a qualitative RG-flow diagram, which is valid for $d>d_L\simeq 8.066$ (The possibility of such RG flow was first raised in \cite{BM-Old}). Note that since MK-RG is functional for disordered systems, one cannot show the full RG flow. The representation in fig. \ref{Fig:diagramSGH} corresponds to its projection on two variables only. The variable we choose are  $\frac{T}{V_J}$ and $\frac{V_H}{V_J}$, as usually done for the RFIM,  with which it shows analogies.
 When the field is not present (on the x-axis) we can find the usual (and already well studied) transition between paramagnet and zero-field SG \cite{MKSG}. This transition is characterized by a finite temperature 
 critical fixed point $\frac{T_c}{V_J}$ that persists down to $d=d_L^{SG}=2.58$ (value that compares well with results of numerical simulations \cite{dLboettcher}). For $V_H=0$ and $T<T_c$ the system is attracted towards the zero-temperature spin glass without field fixed point 
 that we have called SG. This FP is characterized by the coupling growing as $V_J^{(n)}\propto \ell^{\theta_0}$, where $\ell$ is the renormalisation length reached at step $n$: $\ell=2^n$ and
 the exponent $\theta_0$ is dependent on $d$. 
 If the external field is not present, it is not generated under renormalisation. However the SG FP is not stable under field perturbation: even an infinitesimal field is renormalised as
 $V_H^{(n)}\propto \ell^{\frac{d}{2}}$, leading away from the SG FP. The reason is that  $\theta_0<\frac{d}{2}$, as found explicitly within MK-RG and predicted by DT. Hence, even an infinitesimal field always becomes as strong as the coupling after a certain step of renormalisation and leads to the
 instability of the SG fixed point. \\
%On this basis DT wrongly concluded that there can not be SG phase in field.
 One should not conclude, however, that the presence of the field necessarily destroys the SG phase. In fact, while for $d<d_L$ the field continue to grow faster than the coupling and the flow is attracted by the paramagnetic (PM) fixed point 
 $(\frac{T}{V_J}-\frac{V_H}{V_J})=(\infty-\infty)$, for $d\ge d_L$ the flow is the one in Fig. \ref{Fig:diagramSGH}: an infinitesimal field switched in top of the SG-FP
 grows faster than the coupling under RG but when the ratio $V_H/V_J$ reaches a certain finite value, the variances change behaviour. The field and the coupling start
 growing at the same speed: $V_H^{(n)}\propto V_J^{(n)}\propto \ell^{\theta}$, $\theta<\theta_0$. The flow reaches a new zero temperature fixed point 
 $(T/V_J,v_{\overrightarrow{h}}/V_J)=(0,(v_{\overrightarrow{h}}/V_J)^*)$, that we call SGH in Fig. \ref{Fig:diagramSGH}.
 This is the FP governing the spin-glass phase in the presence of field. It is thus different from the one in zero field. This is the first new and very useful 
 information that we get from MK RG for the spin glass in the field: It is not sufficient to demonstrate that the zero field fixed point is unstable under the introduction of an infinitesimal field to state that there not exist a SG phase in a field. Actually there can be two different fixed point in presence and absence of field, as we found.\\
The other important result is that for $d\ge d_L$ there is a critical line in the $\frac{T}{V_J}-\frac{V_H}{V_J}$ plane, the AT line, dividing the paramagnetic phase from the spin-glass one. It starts at zero field at the critical temperature
 and it ends on a new zero temperature critical fixed point that we called $\text{SGH}_c$. It is stable on the direction identified by the critical line and unstable in the other one.
 The field and coupling grow at the critical point as $V_H^{(n)}\propto V_J^{(n)}\propto \ell^{\theta_U}$, $\theta_U<\theta$. Both $\theta$ and $\theta_U$ are dependent on $d$.\\
 This implies that also the critical point is described by a zero temperature FP. Note that 
 a FP is usually called a "zero-temperature" one when the energy scales (couplings and fields)
 become much larger than the temperature. The terminology fixed point and critical point should not be confused: the former refers to the FP of the RG flow whereas the latter to the 
 location in the phase diagram where the phase transition takes place. Hence, a zero-temperature FP can be associated to a finite temperature phase transition, as it is the case for the RFIM \cite{T0_PT,MKrfim}, 
 and as we found for the transition of SG in a field for $d\ge d_L$.\\
The fact that the critical FP is a zero temperature one has deep consequences: a critical FP is identified by three independent exponents, instead of two for usual phase transition \cite{T0_PT}.
 The third one is the $\theta$ exponent that we already introduced. This is connected to the fact that two different correlation functions can be identified, the connected one, associated to 
 thermal fluctuations:
 \begin{equation}
 G_{c}(r)=\overline{(\<\sigma_0\sigma_r\>-\<\sigma_0\>\<\sigma_r\>)^2}=\frac{T^2}{r^{d-2+\eta}}g(r/\xi)
\end{equation}
and the disconnected one, associated to disorder fluctuations:
\begin{equation}
 G_{d}(r)=\overline{\<\sigma_0\>^2\<\sigma_r\>^2}-\overline{\<\sigma_0\>^2}\cdot\overline{\<\sigma_r\>^2}=
 \frac{1}{r^{d-4+\tilde{\eta}}}g_{dis}(r/\xi)\,.
\end{equation} 
 
Their behaviour is associated to different critical exponents, $\eta$ and $\tilde{\eta}$, linked by the relation $\tilde{\eta}-\eta=2-\theta$.
In usual MF theory for SG in the field, all the correlation functions that one can create, included the connected and disconnected ones, has the same behaviour 
because they are all dominated by the only critical eigenvalue of the Hessian, the so called replicon \cite{DeDomGiard}. 
The low-temperature phase identified by the MK RG is thus different from the MF RSB one. It is also different from what advocated by DT, for which there is no transition in field.
The MK RG thus predicts the existence of a completely new phase. 
This call for new complementary investigations. We shall discuss perspectives and possible issues in the conclusion.\\
In the following section we present an analysis of the MK-RG
eqs. in the limit of infinite dimensions. Although the limit of infinite dimensions is not expected to be correctly captured by MK-RG, this analysis is instructive since the MK-RG eqs. can be solved analytically in this case.

\subsection{The $d\rightarrow \infty$ limit} \label{sec:d_inf_SGH}

In ref. \cite{gardner} the limit $d\rightarrow \infty$ ($p\rightarrow \infty$) was studied in the case of zero field. Applying the central limit theorem to the MK-RG equations, 
one obtains that in the $d\rightarrow \infty$ ($p\rightarrow \infty$)  the distribution of the couplings becomes Gaussian, thus the whole flow is fully characterized by the evolution of the variance $V_J^{(n)}$ (the mean is zero), for which we can find the
exact FP of the iteration. Thus the RG is not functional but one has to keep track of one variable only.\\
When a field is introduced, it modifies the equation for $V_J^{(n)}$ \cite{SupplSGH}. The distributions of couplings and fields are both Gaussian, 
and we will have coupled equations for $V_H$ and $V_J$. We can compute the correlation $<V_H^{(n)} V_J^{(n)}>$ that will depend on their initial value. If fields and couplings are uncorrelated,
they will remain uncorrelated during the iteration.\\
When the equations for $V_J$ and $V_H$ are analyzed in the limit $p\rightarrow \infty$, 
we can see that as long as an infinitesimal field is present, it will be renormalised to become large, as found previously. The coupling grows for sufficiently small temperatures and small fields. Since also the critical point is described by a zero-temperature FP the equations 
can be simplified by focusing on the limit $\beta J,\beta h\gg p$.
In this case eqs. (\ref{eq:Jn_SGH}) and (\ref{eq:Hn_SGH})  become:
\begin{align}
\nonumber
J^R_{12}=&\frac{1}{4}\[[
%|J^p_{13}+J^p_{23}+H^p|+|J^p_{13}+J^p_{23}-H^p|
%-|J^p_{13}-J^p_{23}+H^p|-|-J^p_{13}+J^p_{23}+H^p|+\right.\\
\sum_{i=1}^{p}\((|J^i_{13}+J^i_{23}+H^i|+|J^
i_{13}+J^i_{23}-H^i|
-|J^i_{13}-J^i_{23}+H^i|-|-J^i_{13}+J^i_{23}+H^i|\))\]]\\
%\end{align}
%\begin{align}
\overrightarrow{h}^R_1=&\overrightarrow{h}_1^p+\frac{1}{4}\[[
%|J^p_{13}+J^p_{23}+H^p|+|J^p_{13}-J^p_{23}-H^p|
%-|J^p_{13}+J^p_{23}-H^p|-|-J^p_{13}+J^p_{23}+H^p|+\right.\\
\sum_{i=1}^{p}\((|J^i_{13}+J^i_{23}+H^i|+|J^i_{13}-J^i_{23}+H^i|
-|J^i_{13}+J^i_{23}-H^i|-|-J^i_{13}+J^i_{23}+H^i|\))\]]
\label{Eq:T0}
\end{align}
Starting from eq. (\ref{Eq:T0}), one can obtain the following recursion relation on the variable 
$x=\left(\frac{v_{\overrightarrow{h}}}{V_J}\right)^2$:
\scriptsize
\begin{align}
\nonumber &x^{n+1}=F(x^n)\equiv\frac{f_1(x^n)}{f_2(x^n)}\\
&=\frac{\iiint_{-\infty}^{\infty}dJ_1dJ_2dhe^{-h^2/2}e^{-J_1^2/2}e^{-J_2^2/2}
\((|J_1+J_2+2\sqrt{x^n}h|+|J_1-J_2+2\sqrt{x^n}h|
-|J_1+J_2-2\sqrt{x^n}h|-|-J_1+J_2+2\sqrt{x^n}h|\))^2}
{\iiint_{-\infty}^{\infty}dJ_1dJ_2dhe^{-h^2/2}e^{-J_1^2/2}e^{-J_2^2/2}
\((|J_1+J_2+2\sqrt{x^n}h|+|J_1+J_2-2\sqrt{x^n}h|
-|J_1-J_2+2\sqrt{x^n}h|-|-J_1+J_2+2\sqrt{x^n}h|\))^2}.
\label{Eq:F(x)}
\end{align}
\normalsize
We ignored the term coming from the $p$-th branch on which we do not move the fields because it is sub-leading when $p\rightarrow\infty$.
Eq. (\ref{Eq:F(x)}) can be iterated numerically, two fixed points can be identified: the zero field one, $x=0$, that is unstable as expected,
and a second, attractive one, around $x^*\simeq(5.045)^2$.
This value is compatible with the limit $p\rightarrow\infty$ of the numerical results at finite $p$.
Being the equation for $V_J^{(n+1)}$:
$$(v^2_J)^{(n+1)}=(v^2_J)^{(n)}\cdot p \cdot f_2(x^n)=(v^2_J)^{(n)}\cdot 2^{2\theta},$$ 
the exponent $\theta$ of the stable fixed point can be found as: $\theta=\frac{d-1+\log_2 (f_2(x^*))}{2}$.
This leads to $\theta_{\text{SGH}}(d)\simeq(d-1)/2-2.425$, which is actually in good agreement with the numerical values found for $d>8$.\\
The third, unstable critical fixed point $\text{SGH}_c$, the one associate to the transition, can not be found in the limit $d\rightarrow \infty$ by using the central limit theorem.
In fact with the chosen scaling of the Hamiltonian, this fixed point goes as $x_c\propto p$, and in this limit
the leading order in the coupling is 0 (as can be seen from eq. (\ref{Eq:T0})). One thus should go beyond the leading order (and beyond the central limit theorem)
to find it. We can nevertheless obtain its properties from the numerical iteration at finite but large $p$. We find that $\theta_U(\infty)=0$ for $p\rightarrow \infty$. Hence,  
in the $d=\infty$ limit the transition found from MK-RG becomes standard (no more a zero temperature one). Moreover, due to the fact that $x_c\propto p$, the field needs to be rescaled by a factor $\frac{1}{p}$ to have a well defined energy. In this way the other two fixed points SG and SGH collapse
together on this scale. The phase diagram of the 
infinite dimensional limit is thus quite  different from the finite dimensional counterpart.

\section{Disordered Models of Glasses}
In this section we will first review the state of the art, present the results obtained for the disordered models of glasses by MK renormalisation and discuss the new perspectives they offer. 

\subsection{State of the art}\label{sec:intro_glass}
If for spin-glasses there are debates on the nature of the phase transition, for glasses
there are even debates on the existence of the transition itself which, depending on the groups, 
is proposed to be a phase transition at zero-temperature \cite{GarrahanChandlerreview}, at finite temperature \cite{RFOT}, avoided \cite{kivelsontarjus} or even not a phase transition at all 
\cite{dyre}. 
See \cite{berthierbiroli} for a review. \\
Super-cooled liquids are characterized by a viscosity and an equilibration time that grow in a very fast way lowering the temperature. The glass transition seen in experiments corresponds to the point at which the relaxation time is so large that for all practical purposes the 
liquid does not flow anymore and it has become a rigid amorphous solid. Understanding 
the physical mechanism behind the huge growth of the relaxation time---more than 14 decades in a quite restricted temperature range---is the problem of the glass transition.    \\
One of the most famous and promising theories for glasses is the so called Random First Order Transition (RFOT) theory \cite{RFOT} that predicts a real transition at a finite temperature. 
The RFOT theory originated from the observation that the glassy phenomenology is similar to the one of some disordered MF spin models, which can be exactly solved by the so called 
1step Replica Symmetry Breaking (1RSB) theory and display a thermodynamical transition at a certain temperature $T_K$. The simplest example is the Random Energy Model (REM). 
At $T_K$ an infinite number of amorphous states emerges through a bona fide phase transition. Within RFOT theory the phenomenon of the glass transition observed in 
so many different systems is therefore explained as the approach to a true phase transition of a new kind. 
Disordered MF systems and real structural glasses are however deeply different: the former live in infinite dimensions, have discrete degrees of freedom and the Hamiltonian contains
explicit disorder. Contrary, the latter live in three dimensions, the degrees of freedom are the position of real particles, thus they are continuous, and the Hamiltonian does not contain
explicit disorder but the disorder is self-induced. The great breakthrough of Kirkpatrick, Thirumalai and Wolynes was to propose that despite these differences these systems 
share the same statistical properties of the energy landscape. This was first suggested in the
seminal paper on density functionl theory by Singh, Stoessel and Wolynes \cite{DFT} and 
very recently shown to hold by the exact solution of hard spheres in the limit of infinite dimensions \cite{HSHighd}.\\
Despite a lot of successful prediction of the MF theory of the glass transition, there are clearly effects that can be captured only going beyond MF.  In fact as soon as $d\neq \infty$ the physics change drastically: metastable states,
that are the key ingredient in the glassy state in $d=\infty$, have no more an infinite life-time. In finite dimensions the system is able to cross barriers between different metastable states,
while this phenomenon is prohibited in $d=\infty$, where the barriers are infinitely high. 
Glassy dynamics seen in experiments cannot therefore be captured within MF theory and
is encoded in non-perturbative effects in $1/d$ related to growing time and length scales \cite{RFOT}. With the aim of describing this regime, a scaling theory was put forward by Kirkpatrick Thirumalai and Wolynes from
the very beginning and a lot of recent results brought interesting new information.

Once again MK-RG, that is a non-perturbative method, can be of great use in this situation: it
 has the potentiality of shedding light on the role of finite dimensional non-perturbative fluctuations, as well as giving a first description of the RG flow and FPs responsible for the
 glass transition.  In the following we shall present our recent results obtained focusing 
 on the disordered models of glasses that present a RFOT transition at the mean-field level. 
 This represents a first step toward an RG theory of the glass transition.

\subsection{The MK-RG flow for disordered glass models}
We now present the results obtained applying MK-RG  to the disordered glass models defined previously. As for SGs in a field, there exist a lower critical dimension below which no transition
is present. We now focus on $d\ge d_L$ and discuss later what happens for $d<d_L$. \\
The renormalisation eq. (\ref{eq:RG1}), with the movement of the fields for the model with Hamiltonian (\ref{eq:H_Mvalue}) explicitly reads
\begin{align}
\nonumber
-\beta E^R_{1,2}(\sigma_1,\sigma_2)=&\\
=&\ln\[[\sum_{\sigma_3^p=\pm1}
\nonumber
e^{-\beta \((E_{1,3}(\sigma_1,\sigma_3)+E_{3,2}(\sigma_3,\sigma_2)\))}\]]+\\
&+\sum_{i=1}^{p-1}\ln\[[\sum_{\sigma_3^i=\pm1}
e^{-\beta \((\overrightarrow{E}_{1,3}(\sigma_1,\sigma_3)+\overleftarrow{E}_{3,2}(\sigma_3,\sigma_2)\))}\]]
\label{Eq:RGiterGlass}
\end{align}
where we defined $$\overrightarrow{E}(\sigma,\tau)=E(\sigma,\tau)+H_L(\sigma)-H_L(\tau)-C=-J(\sigma,\tau)-H_L(\tau)-H_R(\tau)$$
and $$\overleftarrow{E}(\sigma,\tau)=E(\sigma,\tau)-H_R(\sigma)+H_R(\tau)-C=-J(\sigma,\tau)-H_L(\sigma)-H_L(\tau).$$

\begin{figure}[t]
 \includegraphics[width=\textwidth]{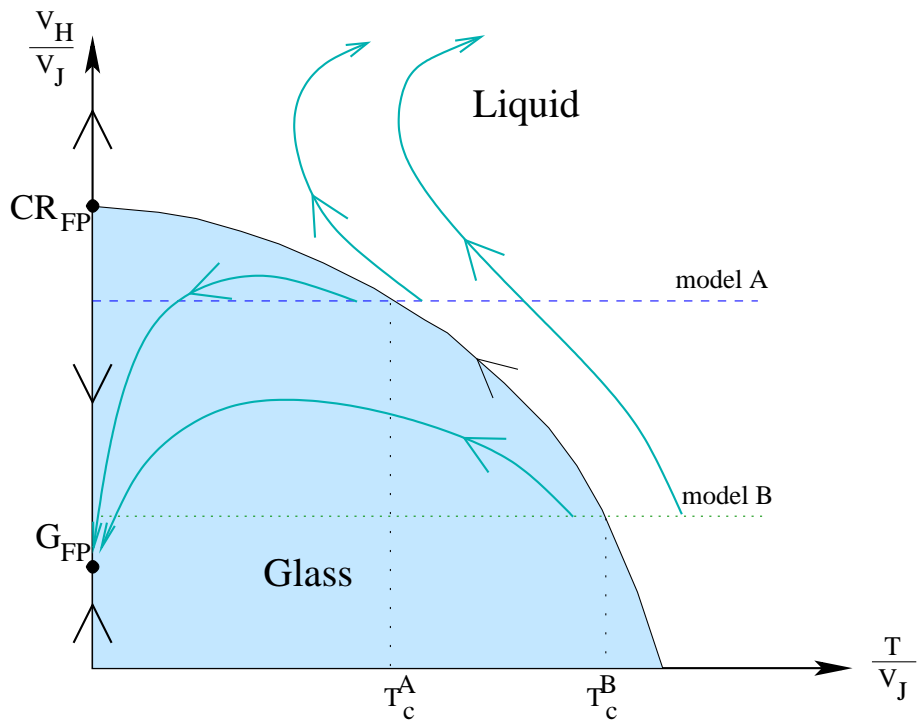}
\caption{Qualitative RG diagram for different glassy models $d\ge d_L(M)$. Two zero temperature FP are present, the critical one, and the glassy one (from ref. \cite{MKREM}).}
\label{Fig:diagramGlass}
 \end{figure}

When the MK RG, as described in sec. \ref{Sec:MKRG}, is applied to models in eq. (\ref{eq:H_Mvalue}), one finds a phase transition and 
the RG flow sketched in Fig. \ref{Fig:diagramGlass} for $d\ge d_L(M)$ (as before we present its projection on the plane $V_H/V_J$ and $T/V_J$ where $V_H$ and $V_J$ are the variances
of the fields and the couplings \footnote{Note that by symmetry the variance of $H(\sigma)$ does not depend on $\sigma$ and likewise for $J(\sigma,\tau)$.}).
The analogy with Fig. \ref{Fig:diagramSGH} - the one for SGs in a field - is evident: 
the low temperature phase is controlled by a zero temperature fixed point that we called $\text{G}_\text{FP}$ and the critical FP $\text{CR}_\text{FP}$ is once again a zero-temperature one.

\begin{figure}[t]
 \includegraphics[width=\textwidth]{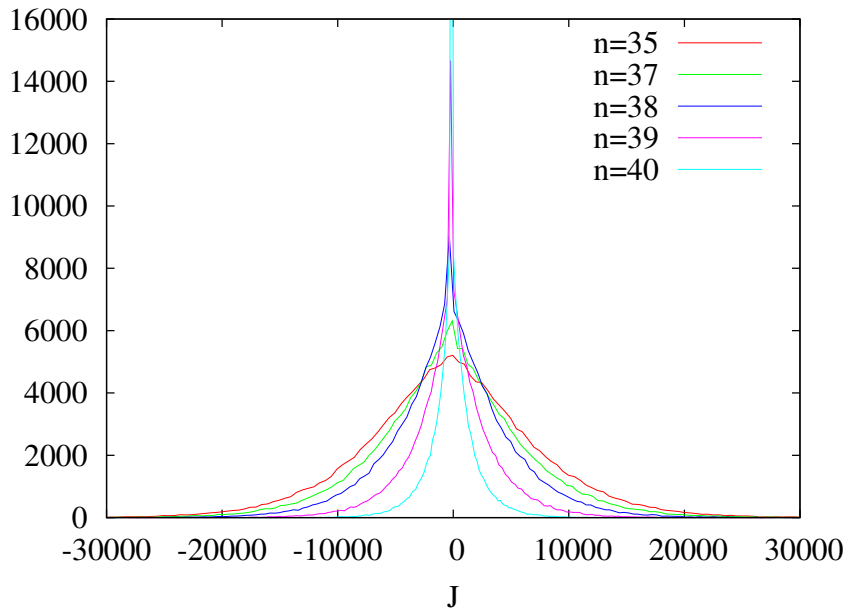}
 \vspace{.7cm}
\caption{Coupling distribution at the $n$-th step of renormalisation for a REM with $M=2$ for which the initial field has been adjusted in order to start
just a little above the critical fixed point at zero temperature.}
\label{Fig:istoJ}
 \end{figure}

As in the SG with field case, the variances of the couplings and fields grow under renormalisation
at $\text{G}_\text{FP}$ and $\text{CR}_\text{FP}$ respectively as $V_H^{(n)}\propto V_J^{(n)}\propto \ell^{\theta}$  and $V_H^{(n)}\propto V_J^{(n)}\propto \ell^{\theta_U}$, with $\theta_U<\theta$
and the exponents are dependent on $M$ and $d$ only and not on the particular $M$-model considered, which just acts as an initial condition for the RG flow.  
The main difference with the case of the SG in field is that for the $M$-value models for $M>1$ the zero field fixed point of Fig. \ref{Fig:diagramSGH} is not present.
Indeed this is expected: for $M>1$ the system looses the spin-inversion symmetry at zero field that is responsible for the existence of the $SG$ fixed point for $M=1$.
Thus for $M>1$, even if we impose an original field $V_H=0$, it will be generated under RG.\\
The Hamiltonian in eq. (\ref{eq:H_Mvalue}) is quite general, depending on the choice of the distribution of the link-energies it corresponds to different models. 
We studied in ref. \cite{MKREM} three of them in detail: the finite dimensional REM \cite{REM1D}, the $M-p$ model \cite{Mp} and the SP model \cite{SP}. 
As anticipated, the flow diagram, the value of $d_L$, of the exponents and the position of the two zero temperature FPs are independent on the choice of the model and they depend only on the values of $M$ and $d$. 
Each model is characterized by a different starting value for the field variance, thus leading to different critical temperature $T_c$, as shown pictorially by the two dashed lines in Fig. \ref{Fig:diagramGlass}.\\
Let us now discuss the feature of the FP distributions of the disorder.  
The qualitative results are valid both for the spin-glass in field and for the glass case.
The fields and couplings are uncorrelated for construction, but they cannot  be independent. However we found that under renormalisation they
become rapidly independent. 
%This is one of the reason why different analyzed models (for which fields and couplings are related in different ways) have the same behaviour for
%the same $M$.
The field distribution under renormalisation tends rapidly to a Gaussian distribution in the whole $(\frac{T}{v_J}-\frac{V_H}{V_J})$ plane (it can be seen analytically in the $d=\infty$ limit). 
More caution is needed for the analysis of the couplings distribution.\\
In order to highlight the peculiarity of the distributions we find, let us recall what happens
at a standard FP, as for example the critical one of SGs in the absence of field. If one starts with a Gaussian couplings distribution, the distribution remains a Gaussian and its variance grows if the system is below the critical temperature, while it decreases above the critical temperature. Crossing the critical point results in a changing on the variance and not in the distribution.
Things are different for the transition in a field or for the disordered models of glasses. 
In Fig. \ref{Fig:istoJ} the evolution of the couplings distribution is shown for different steps of renormalisation for a REM with $M=2$
starting just a little above the critical fixed point at zero temperature. The mechanism to go away from the critical point is quite different from the one described before.
The couplings distribution is no more Gaussian but it splits in a continuous part plus a peak at $J=0$. The height of the peak becomes larger and larger going away from the critical point
under renormalisation. It is not the variance but the weight of the continuous part of the distribution to lower approaching the paramagnetic FP.

The existence of a peak plus a continuous part in the coupling distribution indicates that different samples can be really different: 
some of them are completely disconnected while for some others there can still exist large correlations inside.
This behaviour was already seen in previous numerical and analytic works on RFIM or SGs with field \cite{RFIM_Morone}, \cite{SGHnum2}.
A good RG method should be able to capture these differences. MK has this property.

When the system is approaching the FP associated to the low temperature phase, the distribution of the couplings can still be well approximated by a Gaussian.
In the $d\rightarrow\infty$ limit we analytically find this behaviour (see next section): The distributions of the fields are always Gaussian, the distribution of the couplings is Gaussian at the stable FP, 
that we can determine exactly using the central limit theorem, while it is no more Gaussian at the critical FP where we can not use the central limit theorem any more (see sec. \ref{sec:d_inf_SGH}
and \ref{sec:d_inf_glass}).

\subsection{The $d\rightarrow \infty$ limit} \label{sec:d_inf_glass}

We can now generalize the analytic computation of the $d\rightarrow \infty$ limit for the spin glass case of sec \ref{sec:d_inf_SGH} to the $M$-value Hamiltonian.
As in the spin glass case, if we are at sufficiently small field and small temperature, the flow will spontaneously evolve towards $T=0$. This is thus the limit that we will
analyze in this section.
Again in the $d\rightarrow \infty$ limit the renormalised energies will be Gaussian variables, however there are non-trivial correlations inside the energy matrix that we must take into account.
In the spin glass with field case, the fields and couplings are Gaussian variables, and the only important parameters are $V_J$ and $V_H$. 
This is true also in the $M>1$ case, however we need to answer the question: how to extract typical energy matrices $\overrightarrow{E}(\sigma,\tau)$ and $\overleftarrow{E}(\sigma,\tau)$
given $V_J$ and $V_H$?

Using the properties in eq. (\ref{eq:propJH}), we can compute the correlation:
\begin{equation}
 <\overrightarrow{E}(\sigma,\tau)\overrightarrow{E}(\sigma',\tau')>=
 \begin{cases}
  V_J^2+2V_H^2 & \text{ if } \sigma=\sigma', \tau=\tau'\\
  -\frac{V_J^2}{2^M-1}+2V_H^2 & \text{ if } \sigma\neq\sigma', \tau=\tau'\\
  -\frac{1}{2^M-1}\((V_J^2+2V_H^2\)) & \text{ if } \sigma=\sigma', \tau\neq\tau'\\
  \frac{V_J^2}{(2^M-1)^2}-\frac{2V_H^2}{2^M-1} & \text{ if } \sigma\neq\sigma', \tau\neq\tau'\\
 
\end{cases}
\end{equation}
that can be written in a compact way as

\begin{align}
\nonumber
  <\overrightarrow{E}(\sigma,\tau)\overrightarrow{E}(\sigma',\tau')>\equiv \mathbf{M}=&
  A\((\sum_{\sigma,\tau}|\sigma\tau><\sigma\tau|-\frac{1}{2^M}\sum_{\sigma,\tau.\tau'}|\sigma\tau><\sigma\tau'|\))+\\
  \nonumber
  &+C\((\frac{1}{2^M}\sum_{\sigma,\sigma',\tau}|\sigma\tau><\sigma'\tau|-\frac{1}{2^{2M}}\sum_{\sigma,\sigma',\tau,\tau'}|\sigma\tau><\sigma'\tau'|\))=\\
  =&A(\mathbf{I}-\mathbf{P_L})+C(\mathbf{P_R}-|v><v|)
  \label{eq:matrixM}
\end{align}
where we defined $A=\frac{2^{2M}}{(2^M-1)^2}V_J^2$, $C=-\frac{2^{2M}}{2^M-1}(\frac{V_J^2}{2^M-1}-2V_H^2)$, $\mathbf{I}$ the identity matrix, $\mathbf{P_L}$ the projector on
the subspace of the vector $|v_{\sigma}^L>\equiv\frac{1}{2^{\frac{M}{2}}}\sum_{\tau}|\sigma\tau>$, $\mathbf{P_R}$ the projector on
the subspace of the vector $|v_{\tau}^R>\equiv\frac{1}{2^{\frac{M}{2}}}\sum_{\sigma}|\sigma\tau>$, $|v>$ the vector $|v>\equiv\frac{1}{2^{\frac{M}{2}}}\sum_{\sigma}|v_{\sigma}^L>$.
As expected, the matrix $\mathbf{M}$ does not have components on the subspace $\mathbf{P_L}$.
Once we have written the correlation matrix in the form of eq. (\ref{eq:matrixM}), it is easy enough to compute
$$\sqrt{\mathbf{M}}=\sqrt{A}(\mathbf{I}-\mathbf{P_L})+(\sqrt{A+C}-\sqrt{A})(\mathbf{P_R}-|v><v|).$$
From the last equation we extract the form of the energies $\overrightarrow{E}$:
$$\overrightarrow{E}(\sigma,\tau)=\sqrt{A}\((\eta_{\sigma,\tau}-\frac{1}{2^M}\sum_{\tau'}\eta_{\sigma,\tau'}\))+(\sqrt{A+C}-\sqrt{A})
\((\frac{1}{2^M}\sum_{\sigma'}\eta_{\sigma',\tau}-\frac{1}{2^{2M}}\sum_{\sigma',\tau'}\eta_{\sigma',\tau'}\)),$$
where $\eta_{\sigma,\tau}$ are Gaussian random variables of zero mean and unit variance.
With analogous reasoning we can obtain the form of $\overleftarrow{E}$:
$$\overleftarrow{E}(\sigma,\tau)=\sqrt{A}\((\eta_{\sigma,\tau}-\frac{1}{2^M}\sum_{\sigma'}\eta_{\sigma',\tau}\))+(\sqrt{A+C}-\sqrt{A})
\((\frac{1}{2^M}\sum_{\tau'}\eta_{\sigma,\tau'}-\frac{1}{2^{2M}}\sum_{\sigma',\tau'}\eta_{\sigma',\tau'}\)).$$
Once we know how to extract the link energies, we can proceed to the renormalisation at zero temperature:

$$E^R(\sigma,\tau)=\sum_{i=1}^p \text{min}_{\gamma_i}\((\overrightarrow{E}(\sigma,\gamma_i)+\overleftarrow{E}(\gamma_i,\tau)\))\equiv p V_J^2 f_{\sigma,\tau}(x^2)$$
where again $x^2=\frac{v^2_H}{v^2_J}$.
We compute the new variances of the renormalised fields and coupling and the equation for the renormalisation of $x$:
\begin{equation}
(x^2)^{(n+1)}=F((x^2)^{(n)})=\overline{\frac{\((\frac{1}{2^M}\sum_{\sigma}f_{\sigma,\tau}(x^2)-\frac{1}{2^{2M}}\sum_{\sigma,\tau}f_{\sigma,\tau}(x^2)\))^2}
{\((f_{\sigma,\tau}(x^2)-\frac{1}{2^{M}}\sum_{\sigma}f_{\sigma,\tau}(x^2)-\frac{1}{2^M}\sum_{\tau}f_{\sigma,\tau}(x^2)+\frac{1}{2^{2M}}\sum_{\sigma,\tau}f_{\sigma,\tau}(x^2)\))^2}}
\label{eq:FxREM}
\end{equation}

\begin{figure}[t]
 \includegraphics[width=\textwidth]{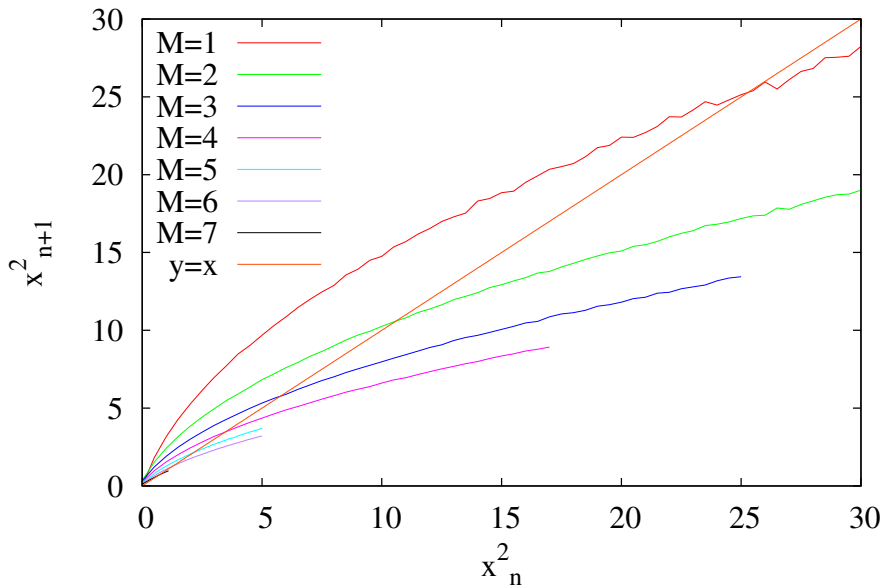}
 \vspace{.7cm}
\caption{$F((x^2)^{(n)})$ as defined in eq. (\ref{eq:FxREM}). A fixed point dependent on $M$ can be identified. For $M=1$ an additional fixed point is found for $x=0$, while it
is not present for $M>1$.}
\label{Fig:REMdInf}
 \end{figure}

The function $F((x^2)^{(n)})$ is shown in Fig. \ref{Fig:REMdInf}.  A fixed point dependent on $M$ can be identified at $x^*(M)\neq0$. 
The value for $M=1$ is in accordance with what found in sec. \ref{sec:d_inf_SGH}.
For $M=1$ an additional fixed point is found for $x=0$ (as found in sec. \ref{sec:d_inf_SGH}), while it is not present for $M>1$.
We can see that $x^*(M)\rightarrow0$ with growing $M$: the FP tends to a zero field FP.
Analogously to what done for the SG in field, we can extract the exponent $\theta$ at the stable FP, that results to be $\theta(M=2)=\frac{d-1-3.7178}{2}$, $\theta(M=3)=\frac{d-1-3.035}{2}$,
that are good predictions for all the numerical MK values found in finite $d$ (just as an example, one could compare the  numerical estimate for $\theta(M=3,d=6.2\simeq d_L(M=3))=1.23714(7)$ 
with the high $d$ prediction $\theta(M=3,d=6.2)=\frac{6.2-1-3.035}{2}=1.0825$).
 
 \subsection{Lower critical dimension and avoided phase transition}\label{sec:d3}
We now focus on $d_L(M)$ and discuss the MK predictions for three dimensional systems. 
The value of $d_L(M)$ as a function of $M$ are presented in the table below:
\begin{table}[h]
 \begin{center}
% \resizebox{8.5cm}{!} {
  \begin{tabular}{| c || c | c | c | c | c | c | c |}
    \hline
  M & 1 & 2 & 3 & 4 & 5 & 6 & 7   \\    \hline 
$d_L(M)$ & 8.07 & 6.93 & 6.21 & 5.70 & 5.32 & 5.00 & 4.7\\ \hline

\end{tabular}
  %  }
\caption{Table of the lower critical dimension $d_L$ as a function of $M$.\label{Tab:dLM}}
\end{center}
\end{table}

The lower critical dimension $d_L$ lowers with growing $M$. The value for $M=1$ corresponds to what found for the case of SG with field.
If an exponential fit is performed on these values, the $M\rightarrow \infty$ extrapolation 
is $$d_L(M=\infty)=4.18\pm0.14$$ Note that large $M$ values (vaguely) correspond to continuous degrees of freedom and hence the large $M$ limit could be representative of realistic interacting particle systems. 
This result predicts that in $D=3$ there is no phase transition for $M$-models and possibly
for realistic glass-formers too.  This of course  has to be taken with a grain of salt since the value for $d_L(\infty)$ will surely depend on the MK approximation, furthermore real particle systems could behave differently than 
$M$-models for large $M$\footnote{Besides having the same mean-field theory, disordered models of glasses and realistic models of supercooled liquids possibly also share the same effective field theory for the overlap field, as first argued in \cite{Stevenson}.}. Nevertheless it is interesting to dwell about what 
these results imply on the physics of three dimensional systems, i.e. what can be learned
on the glass transition observed in experiments from this MK-RG analysis.  \\
In fact, as it is well known, the RG flow is regular in dimension except possibly very close to fixed points. In consequence what happen above $d_L(M)$ is expected to influence the 
behavior also below it. Indeed we find that for $d<d_L(M)$ if the 
temperature is low enough, the couplings and fields start to grow, as if they
were approaching the low-temperature FP. At a certain renormalisation step $n^*$ however, the flow "realizes" that  actually the FP does not exist and starts to deviate towards the PM FP. The value of $n^*$ is larger the more the dimension is near to $d_L$ and it goes to infinity at $d_L$ at the critical temperature.
Defining the correlation length as the length at which $V_J$ reaches half of its initial value, for $d<d_L$ one finds that it grows lowering the temperature but eventually saturates to a finite value at a certain temperature. 
Thus, the existence of a non-trivial FPs above $d_L(M)$ has a direct influence on the physics below, in particular in three dimensions, as it leads to the growing of length-scale.  
The glassy behaviour observed in experiments and numerical simulations is thus explained by the MK RG to be the consequence of an avoided phase transition that exists only at higher dimensions. Note that in $d=3$ for high enough values of $M$ the value of the saturation of the correlation length is so high that experiments and simulations have no hope to distinguish 
it from a true phase transition, the needed size of the systems is simply too large (See Fig. 3 in \cite{MKREM}) 
\begin{figure}[t]
 \includegraphics[width=.5\textwidth]{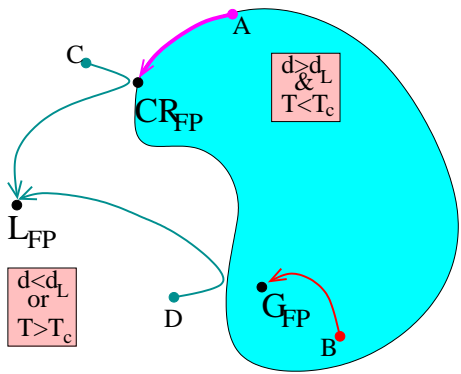}
\caption{The infinite dimensional space of coupling constants and dimensions is 
represented by a 2-dimensional cartoon. There are two stable fixed points,  $\text{G}_{\text{FP}}$ and $\text{L}_{\text{FP}}$, that
correspond respectively to the glass and liquid phase (eventually a third FP corresponding to zero field at $M=1$ can be present). 
The light blue and white regions correspond to the basins of attraction of the two stable FP.
On the boundary between them lies the critical fixed point $CR_{FP}$. 
The initial condition for the flow depends on the temperature, the model and the spatial dimension: for $d\ge d_L$ at $T=T_c $ 
the system lie on the critical manifold and is attracted by $CR_{FP}$ (point A), whereas for $T<T_c$ it starts in the basin of attraction of $G_{FP}$ and falls onto it (point B); 
a model close to criticality, is initially attracted by $CR_{FP}$ but it eventually bifurcates toward $L_{FP}$, this can happen either below or above $d_L$ (point C).
For $d<d_L$ the initial condition always lays in the basin of attraction of $L_{FP}$;
However, if $T<<T_c^{d_L}$ the flow is initially attracted by $G_{FP}$ but it eventually bifurcates toward $L_{FP}$  (point D): 
Due to regularity of the flow equations with respect to spatial dimension, even if $d<d_L$ the system can feel the existence of $G_{FP}$ in higher dimension;
this leads to avoided criticality.}
\label{Fig:artistica}
 \end{figure}
In Fig. \ref{Fig:artistica} a pictorial representation summarizes the properties of the RG flow in the infinite dimensional space of couplings, including dimension of space as an additional axis.
Avoided criticality is represented by the point D: if $d<d_L$ the flow always starts in the basin of attraction of the liquid FP and it will flow towards it.
However, for small initial temperature, the starting point can be close enough to the glassy FP
to feel its attraction in the first RG steps: small size systems would behave as if they should be glassy.

\section{Discussion and Conclusion}
We now recall the main findings presented in this paper and discuss in this context the pros and cons of the MK-RG method for glassy systems. \\
{\bf Zero-temperature fixed points and glassy phase transitions.} The most important result of the MK-RG found both for spin-glasses in a field and disordered models of glasses is that the fixed points associated to the (spin-glass and glass) phase transition are zero temperature ones. 
This means that the characteristic energy of the system, and consequently the energy barriers, 
scale as $\Delta=\ell^{\theta_U}$,
where $\ell$ is the scale over which fluctuations have been integrated out by RG: $\ell=2^x$, with $x$ the renormalisation step. In fact we have shown that this is the scaling of both the couplings and fields variances. 
When the system is in the liquid or paramagnetic phase, it is initially attracted by the critical fixed point, until the static correlation length is reached ($\ell \sim \xi$). At this point
the coupling variance decreases and the system eventually flow towards the high temperature fixed point. 
On scales larger than $\xi$ the system can be considered as an ensemble of weakly
interacting sub-parts (the couplings at that length-scale become small), each one characterised by the energy scale $\Delta(\xi)$.  The time-scale for relaxation should therefore be given
by the Arrhenius law applied to each sub-part, that leads to $\tau\simeq\tau_0 e^{\Delta(\xi)/T}=\tau_0 e^{\Delta_0 \xi^{\theta_U}/T}$ (this standard scaling assumption has been obtained 
recently by non-perturbative dynamical RG method for the Random Field Ising model \cite{TarjusDyn}). We have found that both for spin-glasses in a field and disordered models of glasses the transition exists only above $d_L>3$. Neverteless, 
as explained in Sec. \ref{sec:d3}, even if the glassy fixed point is not present in "low" dimensions, and in particular for $d=3$, the influence of the FP in higher dimensions
leads to growing energies until a saturation length is reached. This implies 
a very strong increase of the relaxation time following an activated dynamics law, 
which is indeed conjectured to hold in different thermodynamic theories of the glass transition \cite{Wbook,kivelsontarjus}. Moreover, it also suggest that a very similar dynamical behavior 
should emerge approaching the glass and the spin-glass transition in a field as indeed found in 
simulations \cite{SGHnum1,SGHnum2} and theoretically conjectured\footnote{The relation between spin-glasses in a field and glasses in three dimensions was first advocated by Moore and collaborators on field theoretical basis and then supported by MK-RG computations. Although their analysis didn't discuss this in terms of avoided fixed point their conclusions are very similar to ours. } in \cite{MKSGHMoore,pspinMoore}.\\
{\bf MK-RG, RSB and many pure states.}
The key ingredient of the MK method applied to disordered systems is the property to be both functional and non perturbative and--at the same time--simple and transparent enough. 
It suffers however of substantial drawbacks. One of the most relevant limitation to study glasses and spin-glasses is the impossibility for MK to identify Replica Symmetry Breaking, if present.
In ref. \cite{gardner} Gardner demonstrated that the low temperature phase for the SG on a hierarchical lattice is always replica symmetric. The reason relies on the finite number of couplings (and consequently of states),
considered by MK-RG. If one renormalises until the RG scale $\ell$ becomes equal to the linear system size then one ends up with two spins characterized by two random fields and a random interaction. Even though the spins can acquire many values in M-value models
these are never enough to capture the complexity of a system displaying a diverging (in the system size) of pure states: indeed, how one can unveil the existence of very many pure states when the number of possible boundary conditions, roughly represented by the two last spins, is finite? Despite this limitation, MK-RG provides some very useful information: the existence of a transition in finite dimensions for the spin-glass in field and for the glass models,
for which the $\mathcal{Z}_2$ inversion symmetry of the spins is not present, automatically suggests an infinite number of pure states for $d\ge d_L$.
In fact having just one pure state is in contrast with the existence of a transition and the existence of a finite number of pure states, even if not impossible, seems very unlikely.
From this point of view, one has to be careful in interpreting MK results for glasses and spin-glasses: they cannot be fully compatible by construction with MF theory, yet as we just discussed, it could be that they reproduce in a very crude way a more complex scenario possibly related to replica symmetry breaking.\\
{\bf Order of the transition: continuous versus random first order}. 
Usually if one wants to understand the continuous or discontinuous nature of the transition in MK RG, it is sufficient to introduce a field coupled to the order parameter of the transition and to study
its behaviour under renormalisation. In the standard notation, we call $x$ the critical exponent associated to this symmetry-breaking field, $\beta$ the exponent associated to the
order parameter, $\nu$ the one associated to the correlation length. Scaling relations imply that $\beta=(d-x)\nu$. Thus if $x<d$, the order parameter is continuous at the transition.
The order parameter of spin-glass and glass transitions should be the overlap between two replicas $a, b$ of the system: $q^{a,b}\propto\frac{1}{N}\sum_i \delta_{\sigma_i^a, \sigma_i^b}$.
Thus one can think to introduce an infinitesimal field $\epsilon$ coupled with $q^{a,b}$ and study its evolution under renormalisation to extract the critical exponent $x$.
We found that  $x<d$ for every value of $M$ and $d\ge d_L(M)$. However $q^{a,b}$ is the average overlap, that is always continuous within mean-field theory (and presumably more generally) at the transition.
The difference between continuos and Random First Order transitions would show up in the full distribution of the overlap $P(q)$. How to capture this within the MK approach is an open problem.\\
{\bf Upper critical dimension and complementary RG approaches.} The other 
important limitation of MK-RG, which exists already for non-disordered models, is the impossibility to capture the upper critical dimension and hence the existence of the
Gaussian fixed point: even in
the ferromagnetic case, the critical exponents do not stick to the MF values for $d\ge4$.
This might not be a serious problem if the relevant fixed point, instead of being Gaussian, is a non-perturbative zero-temperature one, as suggested by MK-RG for spin-glasses in a field and glasses. Nevertheless, a complete analysis should be able to evaluate the competition between these two FPs and their respective basins of attractions. In order to do this, alternative RG methods have to be used. 
For the spin-glass in field we used the Ensemble Renormalisation Group (ERG)
method, which is an approximated way of implementing Dyson hierarchical RG, firstly introduced in ref. \cite{ERG} for the spin-glass without field \cite{MKSGH}.
The ERG method is able to capture (approximatively) the upper critical dimension in the SG without field. By applying ERG to spin-glasses in a field we obtained results in very good agreement with the MK ones, finding a $T=0$ critical FP for $d>d_L\simeq8$ and hence 
strengthening the MK results. It would be very interesting to extend the ERG method to disordered glass models. In any case, as general comment, we stress 
that MK-RG for glassy systems can be a very valuable method but it has necessarily to be supplemented by alternative analysis. A very promising method is the Non-Perturbative Renormalisation Group (NPRG)\`a la Wetterich. However, making it able to capture the complex physics related to zero-temperature fixed points is quite a challenge which for the moment has been only solved in the case of the Random Field Ising model \cite{Wetterich,tarjustissier} and Directed Polymers in Random Media \cite{canet}. A first promising NPRG analysis of spin-glasses in a field \cite{Yaida} has found the existence of a non-perturbative fixed point for the transition, but more general (and tractable) Ansatz for the effective actions able to cope with the physics of zero-temperature fixed points for glassy systems have to be found yet.  \\\\
In this manuscript we have presented and discussed the advantages, the predictive power and also the limitations of MK-RG for glassy systems. The real space RG pioneered by L. Kadanoff long ago for standard phase transition turns out to be a powerful method to investigate also the physics of glassy systems. It has the potentiality of delivering highly non trivial predictions (to "handle with care"
as we underlined). We hope that in the near future field theoretical techniques,
probably based on NPRG, will be developed and used to put on a firmer and more quantitative basis MK-RG results, as it happened for standard critical phenomena. \\\\\\
\noindent
{\bf Acknowledgments}
We thank C. Cammarota, M. Moore, G. Tarjus, P. Urbani for discussions.
We acknowledge support from the ERC grants NPRGGLASS and from the Simons Foundation (N. 454935, Giulio Biroli).


\begin{thebibliography}{100}
\bibitem{spin-glass beyond} M. Mezard, G. Parisi, M. A. Virasoro. "Spin Glass Theory and Beyond`` Eds:M. Mezard et al., World Scientific Press (1987).
\bibitem{droplet} D. S. Fisher and D. A. Huse, J. Phys. A \textbf{20}, L1005 (1987);
D. S. Fisher and D. A. Huse, Phys. Rev. B \textbf{38}, 386 (1988).
\bibitem{dropletBM}  A.J. Bray and M. A. Moore, J. Phys. C \textbf{17}, L463 (1984); 
A.J. Bray and M. A. Moore, ``Scaling theory of the ordered phase
of spin glasses'' in Heidelberg Colloquium on glassy dynamics, ed. by JL van Hemmen and I. Morgenstern, Lecture
notes in Physics vol 275 (1987) Springer Verlag, Heidelberg.
\bibitem{RFOT} T. R. Kirkpatrick, D. Thirumalai, and P. G. Wolynes, Phys. Rev. A \textbf{40}, 1045 (1989).
\bibitem{Wbook} Structural Glasses and Super-Cooled Liquids , Eds: P.G.Wolynes and V. Lubchenko, Wiley (2012).
\bibitem{GarrahanChandlerreview} D. Chandler, J. P. Garrahan, Annu. Rev. Phys. Chem. textbf{61}, 191 (2010).
\bibitem{FRSB} G. Parisi, J. Phys. A \textbf{13},  115 (1980);
J. Phys. A \textbf{13}, 1101 (1980).
\bibitem{Kadanoff}L. P. Kadanoff, Phys. Rev. Lett. \textbf{34}, 1005 (1975).
\bibitem{tarjustissier} G. Tarjus and M. Tissier, Phys. Rev. Lett. \textbf{93}, 267008 (2004);  Phys. Rev. Lett. \textbf{96}, 087202 (2006); Phys. Rev. Lett. \textbf{107}, 041601 (2011).
\bibitem{Migdal}A. A. Migdal, Sov. Phys. JETP \textbf{42}, 743 (1976).
 \bibitem{Berker}A. N. Berker and S. Ostlund, J. Phys. C \textbf{12}, 4961 (1979).
 \bibitem{MKrfim} M. S. Cao and J. Machta, Phys. Rev. B \textbf{48}, 3177 (1993).
 \bibitem{dLboettcher} S. Boettcher, Europhys. Lett. \textbf{67}, 453 (2004); Eur. Phys. J. B \textbf{38}, 83 (2004); 
 Phys. Rev. Lett. \textbf{95}, 197205 (2005).
 \bibitem{Gardner}E. Gardner, J. Physique \textbf{45}, 1755 (1984).
 \bibitem{Antenucci} F. Antenucci, A. Crisanti, L. Leuzzi, Journal of Statistical Physics \textbf{155}, 909 (2014).
\bibitem{MKSGH} M. C. Angelini, G. Biroli, Phys. Rev. Lett. \textbf{114}, 095701 (2015).
\bibitem{MKREM} M. C. Angelini, G. Biroli, arXiv preprint: 1604.03717 (2016).
\bibitem{Stevenson} JD Stevenson, AM Walczak, RW Hall, PG Wolynes, J. Chem. Phys. {\bf 129} 194505 (2008).
\bibitem{derrida}  B. Derrida, Phys. Rev. Lett. \textbf{45}, 79 (1980).
\bibitem{grossmezard} D.J. Gross, M. Mezard Nuclear Physics B \textbf{240},431 (1984).
\bibitem{REM1D} S. Franz, G. Parisi, F. Ricci-Tersenghi, J. Phys. A: Math. Theor. \textbf{41}, 324011 (2008).
\bibitem{Mp} F. Caltagirone, U. Ferrari, L. Leuzzi, G. Parisi, T. Rizzo, Phys. Rev. B \textbf{83}, 104202 (2011).
\bibitem{SP} M. C. Angelini, G. Biroli, Phys. Rev. B \textbf{90}, 220201(R) (2014) 
\bibitem{sompo}  D.J. Gross, O. Kanter, H. Sompolinsky, Phys. Rev. Lett. {\bf 55} 304 (1985).
\bibitem{PermutationMarinari1} E. Marinari, S. Mossa, G. Parisi, Phys.Rev.B \textbf{59}, 8401 (1999).
\bibitem{PermutationMarinari2} L. A. Fernandez et al., Phys. Rev. B \textbf{77}, 104432 (2008).
\bibitem{giap} T. Takahashi, K. Hukushima, Phys. Rev. E 91, 020102 (2015).	
\bibitem{MKSGHMoore} B. Drossel, H. Bokil, M.A. Moore, Phys. Rev. E \textbf{62}, 7690 (2000).
\bibitem{review} E. Marinari, G. Parisi, F. Ricci-Tersenghi, and J.J. Ruiz-Lorenzo in ``Spin Glasses and Random Fields''
, ed: A. P.
Young, World Scientific, Singapore, (1997);
E. Marinari, G. Parisi, F. Ricci-Tersenghi, J. Ruiz-
Lorenzo, F. Zuliani, J. Stat. Phys. \textbf{98}, 973 (2000);
L. Berthier, A.P. Young,  J. Phys.: Condens. Matter {\bf 16}, S729 (2004);
Janus collaboration, J. Stat. Mech. (2010) P06026;
M. A. Moore, H. Bokil and B.Drossel, Phys. Rev. Lett. \textbf{81}, 4252 (1998);
H. Bokil, A. J. Bray, B.Drossel, M. A. Moore, Phys. Rev. Lett. \textbf{82}, 5174 (1999);
H. Bokil, A. J. Bray, B.Drossel, M. A. Moore, Phys. Rev. Lett. \textbf{82}, 5177 (1999);
Barbara Drossel, Hemant Bokil, M. A. Moore, and A. J. Bray, European Physical Journal B \textbf{13}, 369 (2000).
\bibitem{AT} J. R. L. de Almeida and D. J. Thouless, J. Phys. A \textbf{11}, 983 (1978).
\bibitem{SGHnum1}  Janus Collaboration, Phys. Rev. E \textbf{89}, 032140 (2014).
\bibitem{SGHnum2} Janus Collaboration, J. Stat. Mech. P05014 (2014).
\bibitem{SGHnum3} D. Larson, H. G. Katzgraber, M. A. Moore, A. P. Young,
Phys. Rev. B \textbf{87}, 024414 (2013).
\bibitem{BrayRoberts}A. J. Bray and S. A. Roberts, J. Phys. C \textbf{13}, 5405 (1980).
\bibitem{FiniteBasin} A. Moore and A. J. Bray, Phys. Rev. B \textbf{83}, 224408 (2011).
\bibitem{Yaida}P. Charbonneau, S. Yaida, preprint arXiv:1607.04217 (2016).
\bibitem{popdyn} M. M\'{e}zard, G. Parisi, Eur. Phys. J. B \textbf{20} 217-233 (2001).
\bibitem{MKSG} B. W. Southern and A. P. Young, J. Phys. C \textbf{10}, 2179 (1977).
\bibitem{BM-Old} A J Bray and M A Moore,
Journal of Physics C: Solid State Physics {\bf 17} L613 (1984).
\bibitem{T0_PT}A. J. Bray and M. A. Moore, J. Phys. C 18, L927 (1985).
\bibitem{DeDomGiard} C. De Dominicis, I. Giardina, \textit{Random Fields and Spin Glasses}, Cambridge University Press (2010).
\bibitem{gardner} E. Gardner, J. Physique \textbf{45}, 1755 (1984).
\bibitem{SupplSGH} See the Supplementary information of ref. \cite{MKSGH}.
\bibitem{kivelsontarjus} G. Tarjus, S.A. Kivelson, Z. Nussinov and P. Viot, J. Phys. Condensed Matter 17 R1143 (2005).
\bibitem{dyre}  J. Dyre, Rev. Mod. Phys. \textbf{78}, 953 (2006).
\bibitem{DFT} ] Y. Singh,  J. P. Stoessel,  and P. G. Wolynes, Phys. Rev. Lett. \textbf{54}, 1059 (1985).
\bibitem{berthierbiroli} L. Berthier and G. Biroli, Reviews of Modern Physics 83 587 (2011).
\bibitem{HSHighd} J. Kurchan, G. Parisi, F. Zamponi, J. Stat. Mech. (2012) P10012, 
J. Kurchan, G. Parisi, P. Urbani, F. Zamponi, J. Phys. Chem. B \textbf{117}, 12979 (2013), 
P. Charbonneau, J. Kurchan, G. Parisi, P. Urbani, F. Zamponi, J. Stat. Mech. (2014) P10009.
\bibitem{RFIM_Morone} F. Morone, G. Parisi, F. Ricci-Tersenghi, Phys. Rev. B 89, 214202 (2014).
\bibitem{TarjusDyn} I. Balog and G. Tarjus, Phys. Rev. B {\bf 91}, 214201 (2015)
\bibitem{pspinMoore} M.A. Moore, B. Drossel, Phys. Rev. Lett. {\bf 89} 217202 (2002).
\bibitem{ERG} M. C. Angelini, G. Parisi, F. Ricci-Tersenghi, Phys. Rev. B \textbf{87}, 134201, (2013).
\bibitem{Wetterich} J. Berges, N. Tetradis, C. Wetterich, Physics Reports \textbf{363}, 223 (2002).
\bibitem{canet} T. Kloss, L. Canet, N. Wschebor,
Phys. Rev. E {\bf 86}, 051124 (2012).
\end{thebibliography}
\end{document}